\documentclass{article}
\usepackage[utf8]{inputenc}
\usepackage{authblk}
\providecommand{\keywords}[1]{\textbf{\textit{Keywords:}} #1}

\usepackage{amsmath}
\usepackage{graphicx}
\graphicspath{ {./images/} }
\usepackage{hyphenat}
\usepackage{latexsym}
\usepackage{stackrel,amssymb}
\usepackage{esint}

\title{Black holes and the nature of the event horizon}
\author[1,2]{Svetlana Andrusenko}
\author[3,4]{ Daniil Krichevskiy}
\author[4,5]{Valentin Rudenko \thanks{Corresponding Author:  valentin.rudenko@gmail.com}}

\affil[1]{Physics Department, Bauman Moscow State Technical University }
\affil[2]{MIREA — Russian Technological University}
\affil[3]{Institute for Theoretical Physics, University of Bern}
\affil[4]{Sternberg Astronomical Institute, Lomonosov Moscow State University}
\affil[5]{Faculty of Physics, Lomonosov Moscow State University}

\date{2022}

\begin{document}

\maketitle

\begin{abstract}
   The problem of the event horizon in relativistic gravity is discussed.  Singular solutions in general relativity are well known. The Schwarschild metric of a spherical mass is singular at zero ($r = 0$) and at the event horizon ($r = r_g$). Both features reflect the existence of the phenomenon of collapse in general relativity for compact masses exceeding $3M_{\odot }$. A material particle crossing the event horizon falls into a central singularity according to the classical theory of general relativity. In the quantum theory of gravity, there may be no central singularity. The physics of the event horizon is currently being refined. A promising technique is the study of gravitational waves (GW) accompanying the merger of binary black holes at the ringdown stage. GW observations of quasi-normal modes of the newly formed super dense remnant will help clarify the physics of the event horizon.
\end{abstract}


\section*{Introduction}

Formally, Newton's discovery of the law of universal gravitation dates back to 1687 \cite{ Newton}, although the laws of Kepler's celestial mechanics (according to Tycho Brahe's observations) were established much earlier, in the interval 1609-1619 (for literary and historical details, see, for example, \cite{Holton_Brush}). A hundred years after Newton, in 1783, the English priest and amateur astronomer John Mitchell suggested that there are astrophysical objects (bodies) for which the speed of overcoming their gravitational attraction exceeds the speed of light. Fifteen years later, in 1798, the French scientist P. Laplace also expressed and substantiated, using Newton's law of universal gravitation, the idea of the existence of dark bodies in the Universe. At present, it is generally accepted that two scientists were the predictors of dark stars: the Englishman Mitchell and the Frenchman Laplace (More detailed information is given in the work of Cherepashchuk A.M. \cite{Cherepashchuk}). However, in these representations there was no hint of the gravitational collapse of stars, the phenomenon of singularity, the effect of accretion of the surrounding matter and the generation of X-ray luminosity. These details of the physical picture of dark stars, or in modern terminology - Black Holes (BH), - came to science after the creation of the general theory of relativity (GR) and the accelerated development of tools for astronomical observations. The initial dispute about the reality of the existence of BH was resolved in favor of GR predictions about the behavior of matter in strong gravitational fields. 
Experimental search for BHs of stellar masses $(3-100)M_{\odot } $ has been carried out since the mid-60s of the 20th century, and giant BHs, in the centers of galaxies and star clusters $(10^{6} -10^{9}) M_{\odot } $, since the mid-80s. Today these astrophysical objects occupy their equal place among the classical population of the Universe: stars, planets, star clusters, etc. Moreover, when studying the physics and evolution of the latter, it is necessary to take into account the BHs influence. A new field of astrophysical science has emerged – ‘demography of black holes’, which studies their birth, growth and interaction with other objects of the Universe. At present, the number of BHs of stellar and galactic masses is estimated as $10^{7}$ or $0.1\%$ of the baryon mass of the Universe \cite{Cherepashchuk}.
 In parallel with the observations (and stimulated by them), the BH theory was actively developed (as an example, one can point to the well-known monographs \cite{Misner_Torn} or \cite{Frolov1998}) both in the framework of the classical BH and in its generalizations \cite{higher_der}. Note that in a number of alternative theories of gravity, the BH phenomenon is completely absent \cite{RTG}.
 Many amazing properties of BH have been explained. However, due to the complexity of the object, in particular rotating BH, a number of aspects need further clarification. These include the non-trivial concept of "BH event horizon". In this article, special attention is paid to this issue, in fact, its discussion is the main goal of the work.
The structure of the article is as follows. Sections 1 and 2 reproduce the well-known evidence for the inevitability of collapse (the singularity phenomenon) both due to the GR mathematical structure and from the thermodynamical and statistical features of the physics of the evolution of massive stars. Section 3 contains ideas for searching for BH objects among X-ray sources detected by outside atmospheric sensors. Such sources are often binary objects with a BH component, producing X-rays due to the accretion of the surrounding matter onto the BH. This section also describes the recent registration of gravitational radiation from the merger of binary BHs. It has been suggested that the shape of the GW signal should carry information about the conversion of two independent "event horizons" into a third one at the "ring down" stage of the superdense remnant. Section 4 discusses the latest advances in the registration of the BH shadow at the centers of galaxies. Photos of the photon sphere are given allowing one to see the silhouette of the event horizon.

\section{Singular solutions in general relativity}

The general theory of relativity (GR) fundamentally considers situations with strong gravitation fields when the geometry differs significantly from the Euclidean one, which entails the corresponding physical (observational) manifestations (effects).

There are two principal implications of this kind:
\begin{enumerate}
    \item Black holes (BHs) (local geometry change)
    \item Cosmological singularities (the geometry of the Universe as a whole)
\end{enumerate}

An example of a singular solution in GR is the Schwarzschild solution for a spherically symmetric mass $M$:
\begin{equation}\label{eq.1}
    ds^{2}=\left( 1-\frac{r_{g}}{r} \right)  dt^{2} -\left( 1-\frac{r_{g}}{r} \right)^{-1}  dr^{2}-r^{2}\left( d\theta^{2} +sin^{2}\theta \  d\varphi^{2} \right),
\end{equation}
where $(r,\theta,\varphi)$ are spherical coordinates, $t$ is coordinate time corresponding to proper time of an infinitely distant observer and $r_{g}=2GM/c^2$ is Schwarzschild radius (or the gravitational radius) defined with Newton's constant $G$ and speed of light $c$. 

What do these singularities at $r=0$ and $r=r_g$ mean? If we calculate the square of the curvature tensor $R^{ikln}R_{ikln}$ which is an invariant (i.e. doesn't depend on the choice of coordinates) using (\ref{eq.1}) it turns out that it has no singularities at $r=r_{g}$!
This means that the Schwarzschild singularity is a consequence of an unsuccessful choice of the coordinate system.
\subsection{Lemaitre's metric}

Let us consider the Lemaitre reference frame accompanying the particles (dust grains) freely falling along the radial direction towards the center of the field source \cite{Landau}. In this system, the Schwarzschild metric looks like:

\begin{equation}\label{eq.2}
    ds^{2}=d\tau^{2} -\frac{1}{r} dR^{2}-r^{2}\left( d\theta^{2} +sin^{2}\theta \  d\varphi^{2} \right),
\end{equation}
where $R$ is the coordinate of the event (mark of the dust grain), the time $\tau$ is proper time of a particle (taken from the clock of the particle). The function $r\left( \tau ,R\right)$ is given by the equations (for $r_{g}=1$)

\begin{equation}\label{eq.3}
    \begin{cases}\tau -R=\frac{2}{3} r^{3/2}&\text{(expansion)}\\ R-\tau =\frac{2}{3} r^{3/2}&\text{(compression)}\end{cases} \longrightarrow r=\left[ 3/2\left( R-\tau \right)  \right]^{2/3}
\end{equation}
For $r=r_{g}=1$ the singularity is absent. Only the irremovable singularity remains at $r=0\  (R=\tau )$.

Propagation of light signals ($ds^{2}=0$) in the Lemaitre metric along the radial direction ($\left( d\theta^{2} +sin^{2}\theta \  d\varphi^{2} \right)  =0$) is described by the following equation:
\begin{equation}\label{eq.4}
\begin{gathered}
   d\tau^{2} -\frac{1}{r} dR^{2}=0 \Rightarrow  , \\
   \frac{d\tau }{dR} =\pm \frac{1}{\sqrt{r} } \sim \pm \left[ R-\tau \right]^{-1/3}. 
\end{gathered}
\end{equation}

It can be seen that the world line of a particle at rest lies inside the light cone for $r > 1$ and outside the light cone for $r < 1$ \cite{Landau}.

\begin{figure}[ht!]
\centering
\includegraphics[width=.4\textwidth]{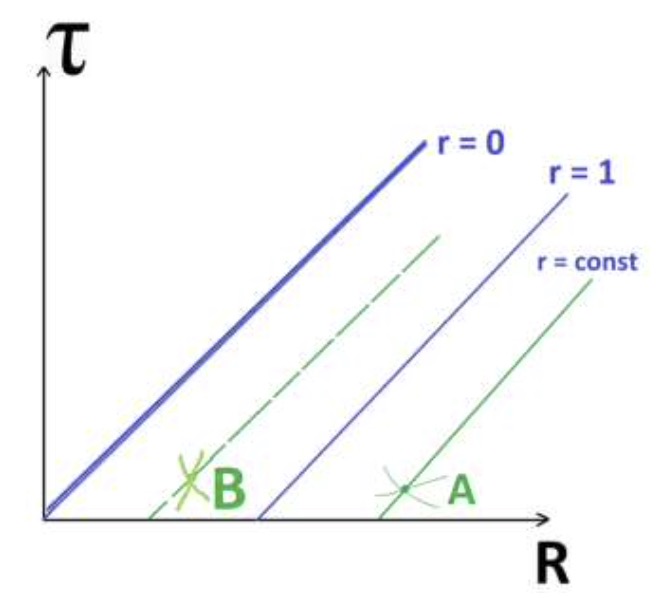}
\caption{Trajectories of particles in Lemaitre coordinates. }
\label{img:01}
\end{figure}

It follows from here that there are no particles at rest beyond the gravitational radius; all particles tend to the center, i.e. collapse. Thus gravitational collapse is one of the fundamental consequences of GR leading to the BHs existences.

\section{Thermodynamics of stars evolution}

Gravitational collapse also follows from thermodynamic and statistical concepts.  Let's consider the initial stage which precedes collapse: the star is a ball of gas held in equilibrium by thermal pressure. Radiation comes at the expense of gravitational energy, with constant compression and, consequently, an increase in temperature in the center. At the advanced stage, the temperature reaches the threshold of nuclear reactions. Then already nuclear energy begins to compensate for the radiation.
At the start of nuclear combustion, the helium fusion reaction mainly takes place: $4p\longrightarrow \ ^{4} He$. When the nuclear fuel runs out, the star turns into a white dwarf in which the gravitational forces are balanced by the pressure of the cold Fermi gas of electrons

\begin{equation}\label{eq.9}
    \epsilon_{F} \ll m_{e}c^{2}. 
\end{equation}

\subsection{Neutron star stage}

When $\epsilon_{F} >\left( m_{n}-m_{p}\right)  c^{2}$ neutronization occurs $e^{-}+p\longrightarrow n+\nu_{e} $. A neutron star of mass $M$ and radus $R$ is formed which does not contain free electrons. 
A simple estimate of the equilibrium condition can be done. The gravity pressure $P_{G}$ 
\begin{equation}\label{eq.10}
    P_{G}R^{2}\simeq \frac{aM^{2}}{R^{2}}
\end{equation}
has to be compensated by the neutron Fermi gas pressure $P_{F}$:
\begin{equation}\label{eq.11}
    P_{F}\sim \epsilon_{F} \sim \frac{1}{m_{n}} n^{5/3}\sim \frac{M^{5/3}}{m^{8/3}} \frac{1}{R^{5}},
\end{equation}
where $n$ is the concentration of the neutron gas. 
The equilibrium condition $P_{G}\simeq P_{F}$ gives
\begin{equation}\label{eq.12}
    \frac{aM^{2}}{R^{4}} =\frac{M^{5/3}}{m^{8/3}} \frac{1}{R^{5}} \cdot const
\end{equation}
From the equilibrium condition, we see that the pressure of the neutron gas increases with compression faster than the gravitational pressure.
Equilibrium radius $R_0$ is
\begin{equation}\label{eq.13}
    R_{0}\approx \frac{c}{4M^{1/3}m^{8/3}} =\frac{1}{\gamma m} \left( \frac{m}{M} \right)^{1/3},
\end{equation}
with $\gamma =Gm^{2}=6\cdot 10^{-39}$. 

If the stars are massive enough, then the neutron gas cannot be cold, it becomes relativistic:
\begin{equation}\label{eq.14.1}
   P_{F} \sim n^{1/3}\sim mc, 
\end{equation}
\begin{equation}\label{eq.15}
    \left( \frac{M}{mR^{3}} \right)^{1/3}  \sim mc
\end{equation}
Then the mass critical value $M_{cr}$ reads
\begin{equation}\label{eq.16}
    M_{cr}=\frac{m}{\gamma^{3/2} } \simeq 1.8\cdot 10^{33}
\end{equation}
When $M>M_{cr}$ the neutron star becomes unstable. For a relativistic neutron gas the pressure is
\begin{equation}\label{eq.17}
P_{n}=\epsilon_{F} \sim n^{1/3}n\longrightarrow P_{n}\sim \frac{M^{4/3}}{m^{4/3}R^{4}}.  
\end{equation}
At the same time, the gravitational mass and the negative pressure of gravity read
\begin{equation}\label{eq.19}
  c^{2}M_{G}\sim \epsilon_{F} \sim n^{1/3}N,  
\end{equation}
\begin{equation}\label{eq.20}
M_{G}\sim \left( \frac{M}{mR^{3}} \right)^{1/3}  \frac{M}{m} \longrightarrow P_{G}=\frac{GM^{2}_{G}}{R^{4}} \sim \frac{1}{R^{6}},
\end{equation}
Thus, the force of gravitational pressure increases faster during contraction than the pressure force of a relativistic Fermi gas, and the star must contract uncontrollably.

\subsection{X-ray sources as BH candidates}

In GR motion under the gravitational radius leads to a fatal collapse. Stars with masses $M>3M_{\odot }$ must collapse at the end of their evolution. What happens at the singularity is unknown. Up to $10^{-33}$ cm, one can use the GR equations, after - a quantum gravity comes into play. 

The awareness of the inevitability of the gravitational collapse of massive and rather compact stellar objects leads to the problem of their experimental (observational) detection. Here are the following key questions.
How to search for BHs? What are their astrophysical manifestations?
Outside the BH, the observer should see the standard Schwarzschild field.
If there is matter in its vicinity in the form of gas or dust then it will be attracted and fall on the BH. As shown in \cite{Shakura_Sunyaev} non-spherical accretion must be accompanied by X-ray emission.
Due to the huge gravitational potential near the BH the speed of matter in the inner parts of the accretion disk is close to the speed of light. Mutual friction and collisions of gas flows in the disk heat up the plasma to temperatures of tens of millions of degrees and form powerful X-rays. The released energy can significantly (by an order of magnitude) exceed the energy of thermonuclear reactions. Due to non-stationary processes in the disk this X-ray emission is variable at times up to  $10^{-3}$ s.

Such a picture is typical for a BH in a binary system with an optical star. When entering the Roche lobe, the BH will pull the matter of the normal component towards itself. Thus, the detection of BHs is associated with the search for astrophysical objects with powerful X-ray luminosity.

\section{Black holes in binary system}

\begin{figure}[ht!]
\centering
\includegraphics[width=.4\textwidth]{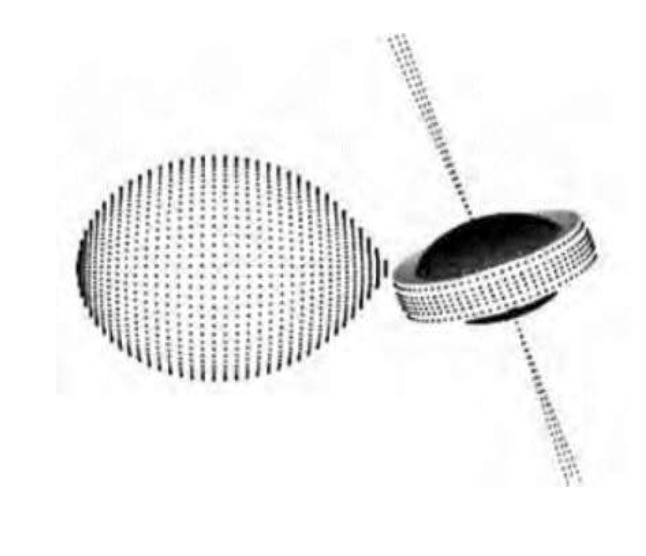}
\caption{Model of an X-ray binary system. Optical star in Roche lobe losing its matter to form the accretion disk around the precessing BH component with relativistic ejections jets. }
\label{img:02}
\end{figure}

\begin{figure}[ht!]
\centering
\includegraphics[width=.7\textwidth]{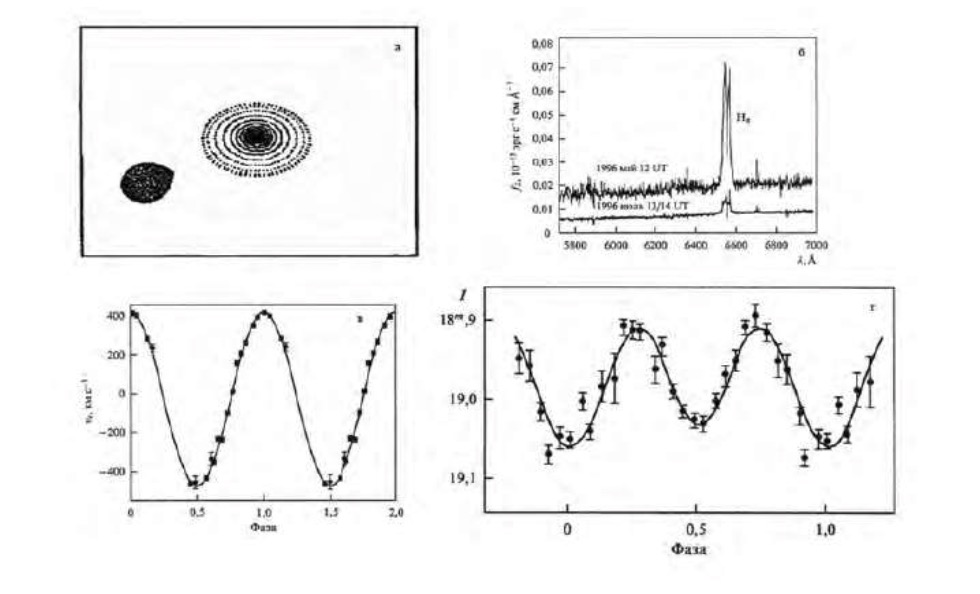}
\caption{Observational data for estimating the mass of a BH in X-ray binary: 
(a)  computer model of the binary system; 
(b)  X-ray spectrum of the binary system ; 
(c)  radial velocity curve of the optical component; 
(d)  light curve of the system (along the y-axis are magnitudes of the object; along the x-axis is the phase of orbital period). 
}
\label{img:03}
\end{figure}

The search for and measurements of the parameters of X-ray sources with BHs has been carried out (and continues) for several decades. Review articles with morphological analysis and systematics of stellar-mass BHs were published, in particular, by A.M. Cherepashchuk in Physics-Uspekhi journal. The main results at the conceptual level are presented in the paper \cite{Cherepashchuk} cited above. It is useful to present a number of illustrations and statistics from this work below (Fig. \ref{img:02} and \ref{img:03}).

\begin{figure}[ht!]
\centering
\includegraphics[width=.6\textwidth]{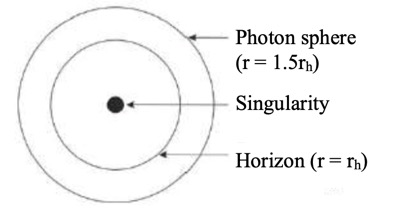}
\caption{Diagram of a Schwarzschild BH in Euclidean space (neighborhood of the event horizon). In the center is the singularity where the matter fell. The singularity is surrounded by a spherical event horizon, with radius $r_g = 2GM/c^{2}$. Next is the photon sphere, in which the captured photons move along closed trajectories, the radius of the photon sphere is $1.5r_g$. Due to the strong curvature of light rays in the BH gravitational field, it acts as a lens for itself. In this case, the size of the event horizon and the photon sphere seem to be increased to an external observer.}
\label{img:04}
\end{figure}

\subsection{Demographics of black holes}

At the moment, thousands of X-ray binaries have been discovered from the boards of orbital X-ray observatories in which NS and BH of stellar masses are detected and studied.
The masses of three dozen stellar BHs in X-ray binaries have been measured, as well as the masses of about 70 NS. 
In this case, the BH masses lie within $(4\div 16)M_{\odot }$ and NS masses -  $(1\div 2)M_{\odot }$ and are concentrated near the value $\sim 1.4M_{\odot }$. 
It was also possible to measure the angular momentum of rotation for more than a dozen BH of stellar masses by comparing their X-ray spectra with theoretical spectra (the model of thin accretion - Novikov and Thorne 1973 \cite{Novikov_Thorne}). The normalized torques of the BH lie in a wide range from $0.98$ to $0.12$, i.e. among stellar BHs, there are both rapidly rotating and slowly rotating. 

\subsection{The physical riddle of the event horizon}

Is this concept related to the presence of a material (observable) surface in collapsars? The study of X-ray binaries showed that their relativistic NS and BH objects differ not only in masses, but also in observational manifestations. In all 70 cases, when a compact object with $M <3M_{\odot } $ shows signs of an observed surface, it is an NS:  X-ray pulsar, radio-pulsar or X-ray burster of the first type requiring the presence of a strong magnetic field. On the contrary, none of the three dozen massive ($M > 3M_{\odot } $) compact objects (candidates for BHs) shows signs of an observable surface in accordance with the GR prediction. Instead, there are only conceivable boundaries - event horizons - light surfaces in spacetime
(which depend on the choice of an appropriate frame of reference). The statistics of relativistic stellar mass objects is already sufficient to accept this significant result - experimental evidence of the absence of observable surfaces in stellar BHs (or the presence of an event horizon): New proof of this has recently been obtained by detecting the gravitational radiation of merging binary BHs.

Four types of BHs are known as solutions to Einstein's equations, in GR. Two of them rotate: Kerr and Kerr-Newman BHs. It is believed that each such BH, losing energy, quickly collapses and becomes a stable BH.

By the no-hair theorem, (except for quantum fluctuations) stable BHs can be completely described at any time by these eleven numbers: mass - energy $M$, momentum $P$ (three components), angular momentum $J$ (three components), radius vector $x$ (three components), electric charge $Q$.
These numbers represent the lingering properties of an object, which can be determined from a distance by studying its electromagnetic and gravitational fields. All other changes in the BH will either go to infinity or be swallowed up by the BH. This is because everything that happens inside (within the BH's event horizon) cannot affect events outside of it.

A falling observer  with increasing acceleration hits a rotating BH in a finite proper time. From the point of view of a distant observer at infinity, it  stops at the event horizon, approaching zero speed relative to a stationary reference point in place, and at the same time rotates around the event horizon infinitely often due to the drag effect of inertial frames of reference. 

Rotating BHs are formed by the gravitational collapse of massive rotating stars or by the collapse of a cluster of stars or gas with a total angular momentum other than zero. Since most stars rotate, it is assumed that most BHs in nature are rotating BHs.

\begin{figure}[ht!]
\centering
\includegraphics[width=.6\textwidth]{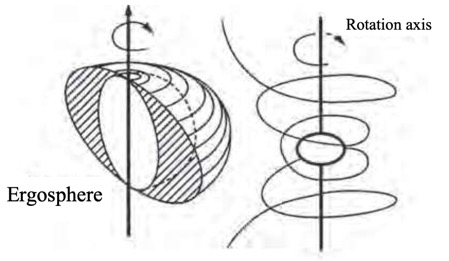}
\caption{Diagram of a Kerr BH. On the left is the event horizon surrounded by an ergosphere in which matter and photons move rapidly entrained by the vortex gravitational field of the BH (Lense-Thirring effect). 
On the right, a beam of light passing near a BH swirls its motion. }
\label{img:05}
\end{figure}

The rotating BHs besides the event horizon have also additional conceivable surface, so called “ergosphere” (Fig. \ref{img:05}). Ergosphere is a region of space-time near a rotating BH, located between the event horizon and the static limit. Objects within the ergosphere inevitably rotate with the black hole due to the Lense-Thirring effect. The characteristic size of a non-rotating BH is given by its gravitational radius $r_g$, and the physical boundary of the BH is the event horizon, a sphere of radius $r_g$. For a rotating BH, the event horizon radius $r_h$ is less than $r_g$, and for an extremely fast rotating BH (the so-called Kerr BH) $r_h = r_g/2$.

At the end of 2006, astronomers gave estimates of the rotation speed. A BH in the Milky Way, GRS 1915+105, can rotate $1.150$ times per second (!), not yet approaching the theoretical upper limit. 
A spinning BH can produce a large amount of energy from its rotational energy. This happens through the Penrose process (vacuum polarization with the birth of real particles and their expansion into BHs and to infinity) in the ergosphere of a BH, a region outside the event horizon. In this case, the rotating BH gradually shrinks down to a Schwarzschild BH, the smallest form from which no energy can be obtained. However, the spin rate of a Kerr BH will never reach zero.

\subsection{GW from coalescing binaries}

\begin{figure}[ht!]
\centering
\includegraphics[width=.6\textwidth]{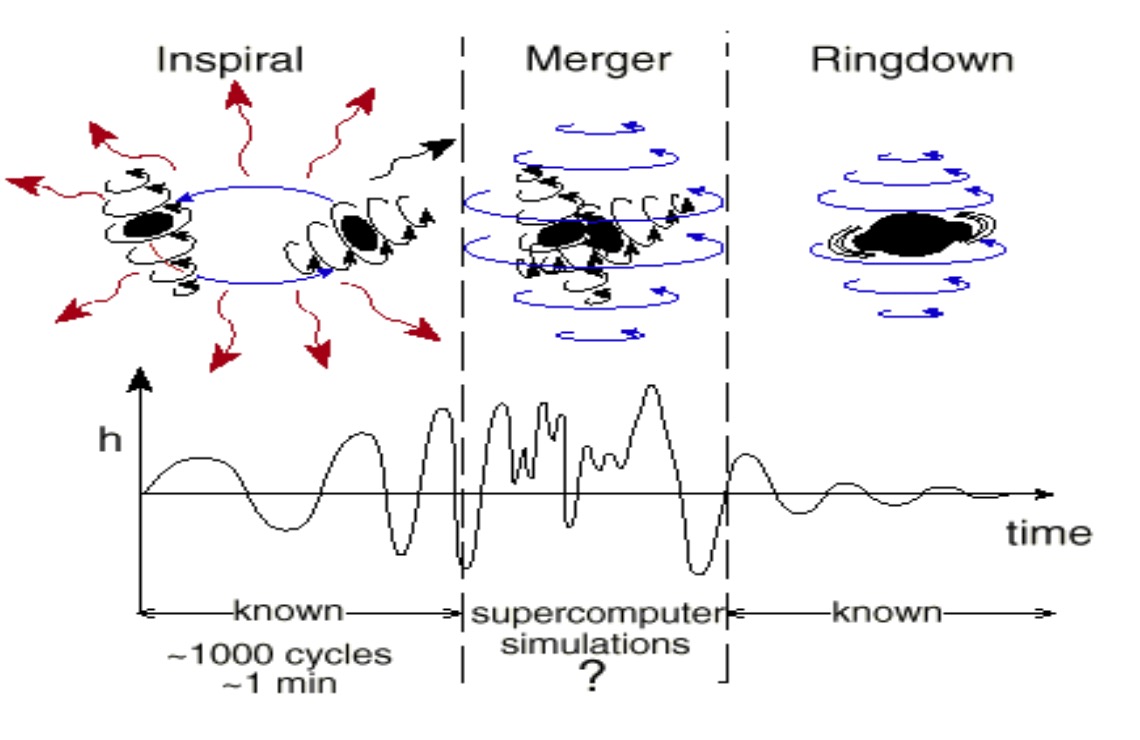}
\caption{Phases of BH binary merger}
\label{img:07}
\end{figure}

\begin{figure}[ht!]
\centering
\includegraphics[width=.9\textwidth]{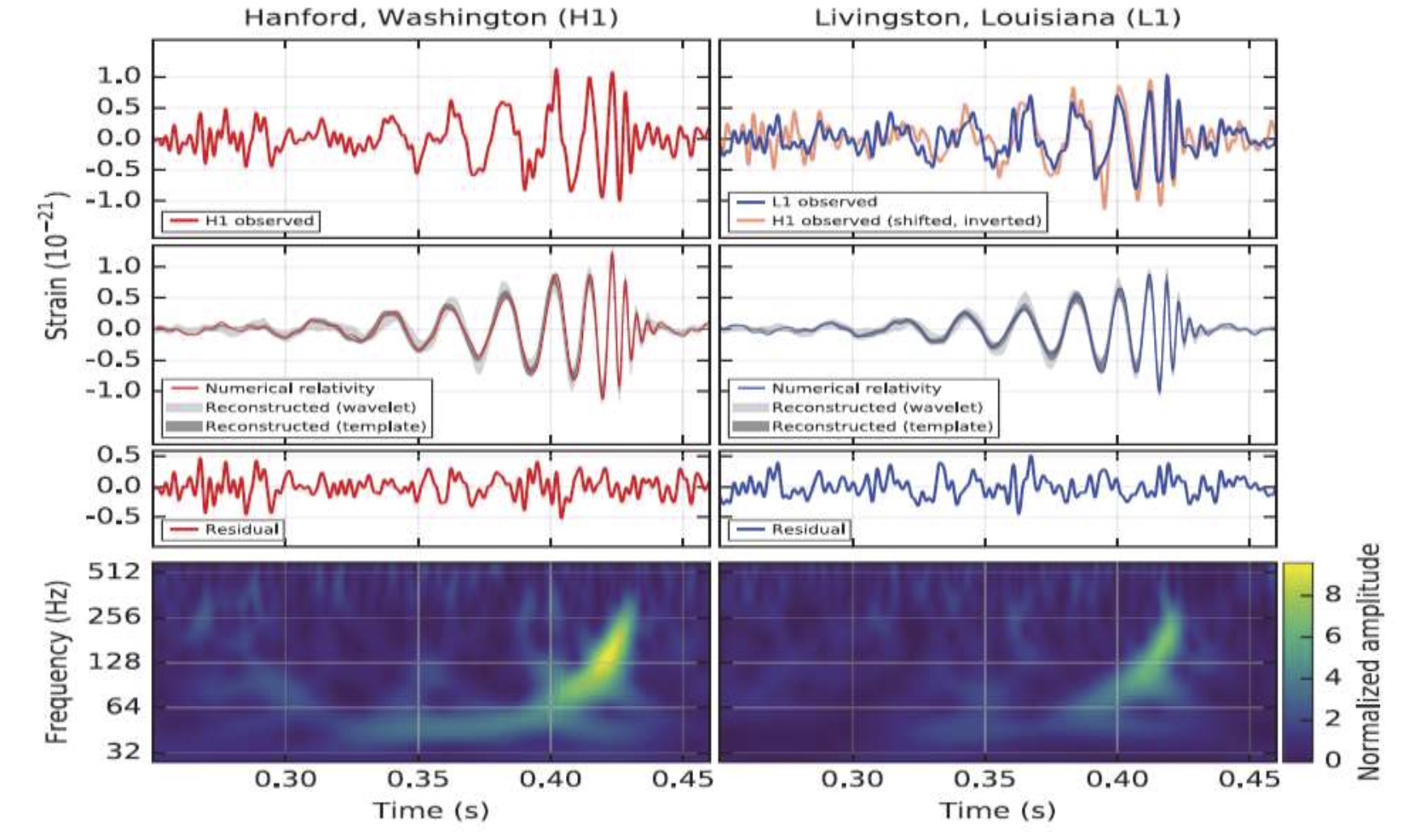}
\caption{Event GW150914 (Sept. 14, 2015 at 09:50:45 UTC). All time series are filtered with a 35–350 Hz band-pass filter and
band-reject filters to remove the strong instrumental spectral lines ;
GW150914 arrived first at L1 and $\sim 6.9 \ ms$ late at H1; (for a visual comparison,the H1 data a real so shown,shifted in time by this amount and inverted to account for the
detectors’ relative orientations);
second row: solid lines show a numerical relativity waveform for a system with parameters consistent with those recovered from GW150914
third row: residuals after subtracting the numerical relativity waveform from the filtered detector time series. bottom row: a time-frequency representation of the strain data, showing the signal frequency increasing over time. (from \cite{first_gw}).   }
\label{img:10}
\end{figure}

A priory the reliable forecast was to search GW radiation signal from the coalescence of compact binary stars (BNS, BBH, NS/BH). For this type of events the signal waveform was accurately modelled at the first stage (in spiral) and last stage (ring down) of the process (Fig. \ref{img:07}). In principle it would allow to use the matched filtering under detection (Fig. \ref{img:10}). 

However much less known was the waveform in the merger phase.  Here a new interesting physics was assumed with the hope of obtaining an increased amplitude of the emitted GWs. In this regard, a lot of effort was expended by various university groups on the calculations of GW in the merger stage. The very first experiments on detecting GW from merging binaries showed that these expectations were untenable. The intensity of gravitational radiation did not increase radically. It is now clear that the maximum amplitude of the GW signal appears in the last orbital cycle of the spiral merger stage. Rate of events depends on the source distance. A few events/year could be accessible to the GW interferometers  LIGO-Virgo.

\begin{figure}[ht!]
\centering
\includegraphics[width=.6\textwidth]{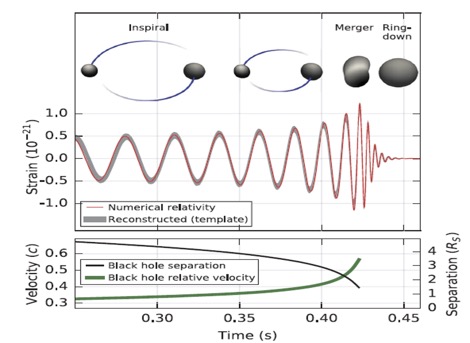}
\caption{Modernized diagram of BH-binary coalescence (from \cite{first_gw}).}
\label{img:09}
\end{figure}

Taking into account the experimental fact (the absence of features of the metric amplitude at the stage of direct contact of the components), the theoretical picture  Fig.\ref{img:07}  of the merger of a relativistic binary should be presented as given in Fig.\ref{img:09}. 

Thus, after the inspiral phase, the ringdown phase immediately follows, corresponding to the so-called. quasi-normal oscillation of the resulting BH. The diagram in Fig.\ref{img:09} also gives insight that the stages of direct contact must contain a "collision of event horizons"

\begin{figure}[ht!]
\centering
\includegraphics[width=.6\textwidth]{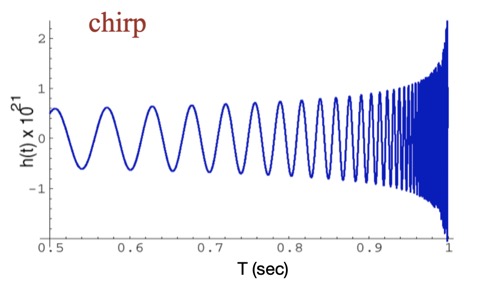}
\caption{Gravitational wave amplitude of a chirp signal.}
\label{img:08}
\end{figure}

GW emitted during the inspiral phase can be described by a chirp signal (see Fig. \ref{img:08} which is defined by so-called chirp mass parameter $\mathcal{M}$: 
\begin{equation}\label{eq.21}
\mathcal{M}=\frac{(m_{1}m_{2})^{3/5}}{(m_{1}+m_{2})^{1/5}} =\frac{c^{3}}{G} \left[ \frac{5}{96} \pi^{-8/3} f^{-11}\dot{f} \right]^{3/5},
\end{equation}
where $m_1$ and $m_2$ are masses of the binary components, $f$ is the chirp frequency and dot denotes the time derivative.

The first ever GW signal was detected in 2015 (GW150914) \cite{first_gw}). It was a merger of two BHs (NS-NS, BH-NS options were rejected) from a distance of $D\sim 400 \ Mpc$. The duration of the signal was $\tau \approx 0.2 \ s$. The signal frequency increased during the inspiral $35 \ Hz \rightarrow 150 \ Hz$. The chirp mass of the binary system was $\mathcal{M}\approx30 M_{\odot } $ with $m_1+m_2\approx70M_{\odot } $ and gravitaional radius  $2GM/c^2\geq 210 \ km$. The signal-to-noise ratio was $SNR \sim 24$ and the dimensionless amplitude - $h\sim 2\times 10^{-21}$.

\subsection{Method of probing the event horizon for a black hole}

One can study the event horizon nature  using the GW observation from coalescence binary BH systems. To prove the existence and study the properties of the event horizon, it is necessary to observe the effects generated by its changes. In case of a binary merger, the resulting BH with a new event horizon is formed from two initial BHs with their own horizons. 
It has to be reflected in the GW waveform at the ringdown stage. Experimentally the task is to detect and study the quasi-normal mode oscillations of the newly formed remnant after the merger i.e. in the process of damping  this oscillations (ringdown phase) \cite{GW_Anderson}.  The study of GW from BH quasi-normal modes is an important test for quantum gravity theories \cite{QNBH}. 

Calculation of the merger dynamics requires numerical simulation (on modern supercomputers) of GR non-linear differential equations. Today such a calculation has been performed for many variants of the two BHs merger. At the final stage, two initial BHs form one new BH with a larger mass, which is in an excited state. Further, these excitations
fade, emitting GW. Some of the emitted GWs are absorbed by the BH itself, and some go to infinity, carrying information about the specific nature of the event horizon. In fact, this is information about the highly non-linear large-scale dynamics of the space-time curvature. An experimental study of these GW signals will make it possible to test the nonlinear equations of GR in all their completeness and adequacy to nature. Apparently, this will provide decisive evidence for the existence of an event horizon for BHs.

\section{Supermassive black holes at the centers of galaxies}

The first estimates of the masses of supermassive BHs in very active galactic nuclei (quasars) were made by Zeldovich and Novikov in 1964 \cite{Zeldovich_Novikov} under the reasonable assumption that the luminosities are close to the critical Eddington luminosity, at which the radiation pressure force balances the gravitational force. It turned out that the masses of these BHs must be very large - hundreds of millions of solar masses.

\begin{figure}[ht!]
\centering
\includegraphics[width=.6\textwidth]{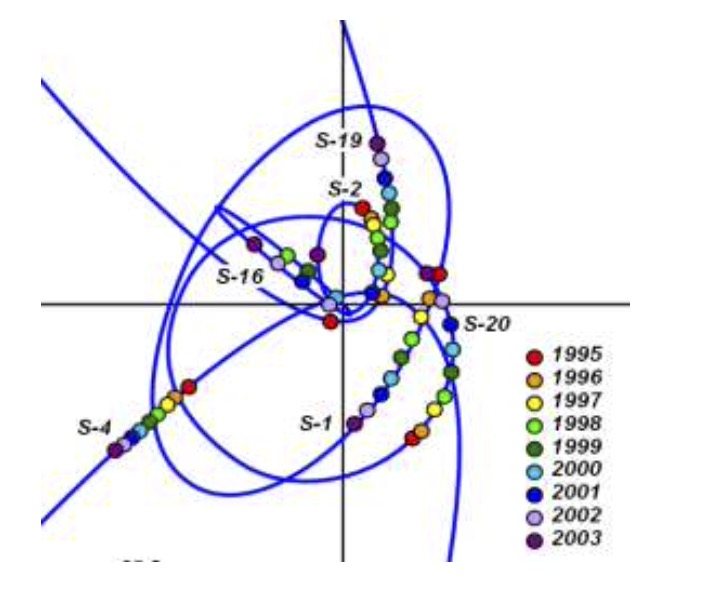}
\caption{Orbits of stars around the center of the Galaxy in the region $r \sim 0.01 \ pc$}
\label{img:11}
\end{figure}

It is now clear that at the centers of most galaxies there are huge supermassive BHs with masses $10^6-10^{10}M_{\odot  }$. To date, reliable methods for estimating the masses of supermassive BHs in the nuclei of galaxies are based on studying the motion of various "test bodies" in the gravitational field of a central supermassive BH (stars, gaseous disks, gas clouds, etc.) (Fig. \ref{img:11}).

In GR isolated BHs cannot be observed due to the infinitely large redshift of photons traveling from the event horizon to a distant observer. However, the dark shadow (silhouette) of a BH (object) can be seen against the background of radiation from matter more distant along the line of the observer lensed by the BH gravitational field. It is also stated there that the minimum size of the shadow can be observed if the same BH is illuminated by photons from the inner part of the accretion disk adjacent to the event horizon.
In general, it is clear that a remotely observed shadow pattern can be
quite bizarre due to the effects of gravitational lensing, the vortex gravitational field of the ergosphere (the Lense-Thirring effect), as well as the Doppler frequency shift of radiation from different zones of the accretion disk.
In particular, recent papers \cite{Dokuchaev}  in the framework of GR and \cite{Bogush} involving a generalization of scalar fields are devoted to the analysis and calculations of the structure of shadows.

\begin{figure}[ht!]
\centering
\includegraphics[width=.6\textwidth]{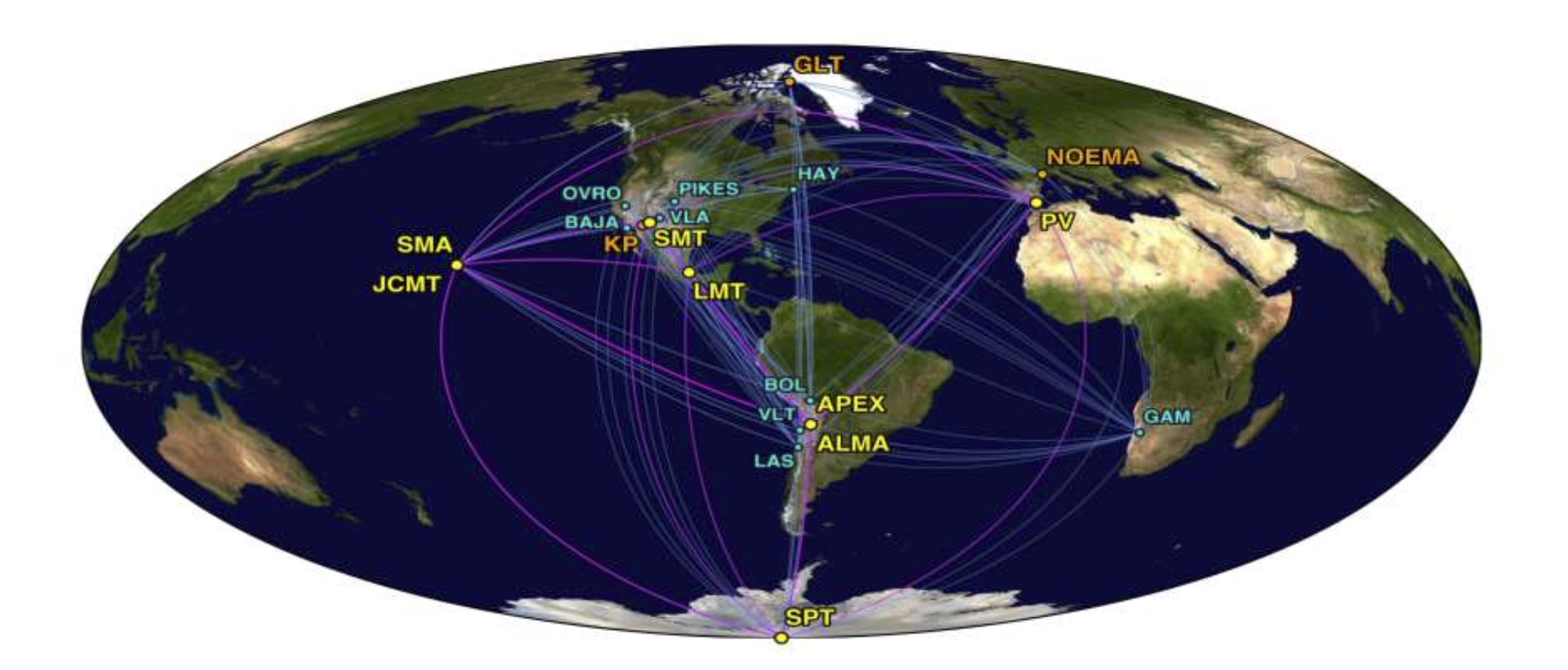}
\caption{The Event Horizon Telescope collaboration. }
\label{img:12}
\end{figure}

The black holes predicted by the general theory of relativity are by their nature (in principle) invisible objects. The event horizon of any black hole is a surface formed by geodesic photon trajectories that do not go to spatial infinity. Direct evidence was first presented in April 2019 by the international collaboration EHT (Event Horizon Telescope) (Fig. \ref{img:12}). The image of the shadow was made by linking 11 radio telescopes on four continents into one huge radio interferometer has obtained the clearest image in history of a supermassive BH in the M87 galaxy. The resolution of the resulting image (several microarcseconds) turned out to be sufficient for the first time to see the shadow of the BH itself, which goes along the event horizon in the center of this object, as well as to estimate the size, plane and luminosity of the accretion disk (substance fall) around it. For a mass of $6\times10^9 M_{\odot }$, the event horizon of this BH has exactly the predicted angular size. The image qualitatively consists of a bright asymmetric ring, interpreted as the glow of an accretion disk, and a dark central part, interpreted as the observed silhouette of a BH.

\begin{figure}[ht!]
\centering
\includegraphics[width=.7\textwidth]{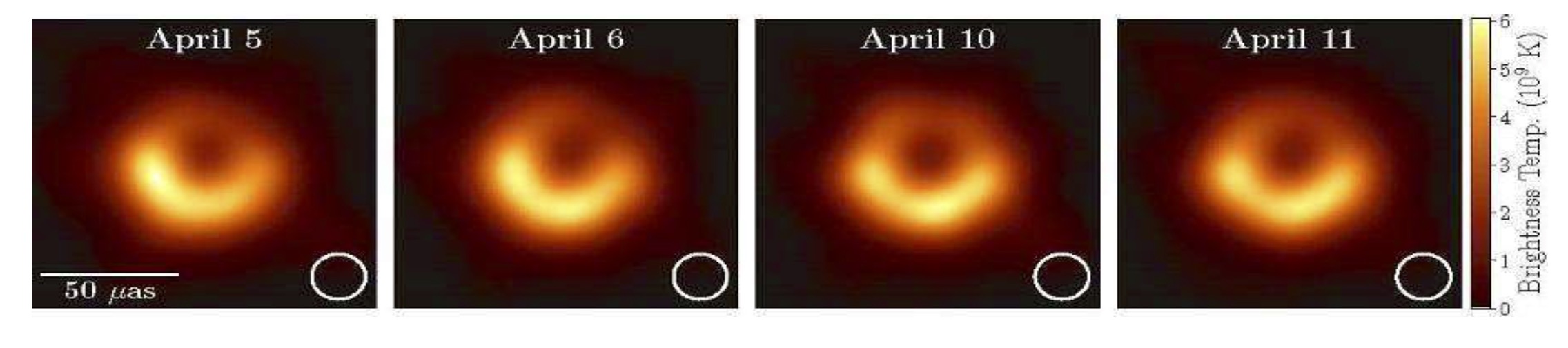}
\caption{EHT images of M87 on four different observing nights (from \cite{EHT_4})}
\label{img:13}
\end{figure}

The object at the center of M87 looks like a BH framed by a glowing crescent - this is radiation from an accretion disk of matter falling into the hole. BHs in M87 were observed by telescopes for four days - April 5, 6, 10 and 11, 2017 so much time in the observation window at eight points on the planet, including the South Pole, there was good weather, which allowed scientists to get a clean picture (Fig. \ref{img:13}) . The image of a BH shown by the EHT collaboration was obtained using its model, in which the hole rotates clockwise. The plasma disk around the hole is almost round - this means that it is almost (but still at an angle) perpendicular to the picture plane in which observers from the Earth look at M87.

\begin{figure}[ht!]
\centering
\includegraphics[width=.6\textwidth]{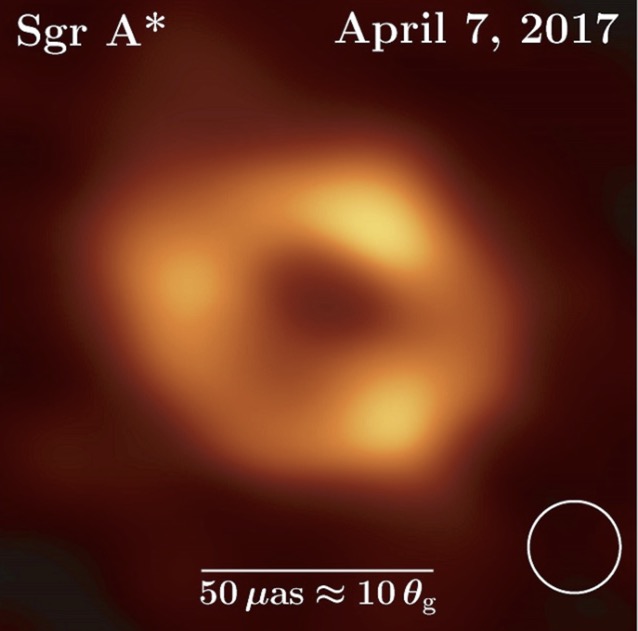}
\caption{EHT image of Sagittarius A* (from \cite{EHT_2022_photo})}
\label{img:14}
\end{figure}

The closest supermassive BH to Earth, Sagittarius A*, is at the center of the Milky Way, $26000$ ly away (while M87 is $55$ million ly away). But its radius is close to the distance between the Sun and Pluto, and it is difficult to see it due to its small angular diameter (its mass is $\sim 4\times10^6 M_{\odot }$ compared to the giant hole in the center of M87 with a mass of $\sim 5\times10^9 M_{\odot  }$).  This is about how to take a picture of a DVD disc lying on the surface of the Moon, or to see the Earth from Alpha Centauri. Despite all the technical difficulties, scientists managed to obtain images of the shadow of a black hole at the center of our galaxy (see Fig. \ref{img:14}), which was announced in 2022 \cite{EHT_2022}.

\section*{Conclusion}

The creation of detectors of gravitational radiation and its successful detection from merging relativistic binary stars opened up a new unique channel for obtaining astrophysical information about the surrounding universe. The development of GW astronomy promises an acceleration in the rate of knowledge of the surrounding world by mankind and the following technical revolution in its development. The list of fundamental scientific problems that the GW astronomy should help to solve already looks impressive. Here we should note the conceptual issues of theoretical physics.

Are the detected GW those as predicted by Einstein's theory? Is the GR theory itself correct? Do the observed BHs correspond to the GR picture? Next, there is a set of questions from astrophysics. What is the nature of observed gamma-ray bursts? And also what is the structure of neutron stars and other compact objects? How to clarify the process of gravitational collapse itself? Finally, there are the inevitable questions of cosmology. What were the initial conditions before the Big Bang? Nature of phase transitions (if any) in the Early Universe ? What is the Dark Energy in reality? Obviously, this list is not exhaustive.

The ability to register and measure the parameters of merging BH through the gravitational-wave channel allows us to significantly advance our understanding of the foundations of the Universe - in cosmology and physics of the microworld. The masses of the merging components can be comparable, but they can also have significant differences. These are events of moderate BH falling onto supermassive BH at the centers of galaxies. In its spiral trajectory, the test mass (smaller BH) will penetrate the zone of strong gravitational field on the last orbits near the event horizon of the supermassive BH. In this case, the structure of the emitted GW pulse should contain unique information about the physics of this still unexplored area. It will be possible to carry out such a research program using space GW interferometers LISA or eLISA when they are created \cite{Rudenko_GW}. The sensitivity of these cosmic detectors should be enough to register BH merger events at the early stages of the development of the Universe, in the so-called epoch of "cosmic noon" from $ z\sim 3 $ to $ z\sim 1.5 $ when the rate of star formation in the Universe, the activity of quasars and galactic nuclei were the highest. In "late cosmos", at $z<1$, space interferometers will be able to continue tracking binary BH merger events with an increased signal-to-noise ratio. It is in this closer epoch that we can expect "spiral signals" from the fall of test BHs on the central supermassive BH.

Today, after the experimental confirmation of the existence of stellar-mass black holes and giant black holes in the centers of galaxies, it seems that GWs can help solve another intriguing mystery of philosophy and natural science. What is a "black hole event horizon"? How much mysticism and reality is here? The way to solve, as the authors of this article think, lies through the accumulation of observations of the shape of the GW radiation accompanying the collisions of binary BHs at the final ringdown stage and its comparison with the prediction given by calculations on damped oscillations of quasi-normal BH modes.

\section{Acknowledgements}
The authors are grateful to their colleagues from Moscow State University, professors Cherepashchuk A.M. and Galtsov D.V. for many helpful discussions. 

\bibliographystyle{ieeetr}

\end{document}